\documentclass[iop,apj,numberedappendix]{emulateapj}
\usepackage{amsmath,amssymb,color}

\begin{document}
\title{Soft X-ray Extended Emissions of Short Gamma-Ray Bursts as \\ Electromagnetic Counterparts of Compact Binary Mergers; \\ Possible Origin and Detectability}

\author{Takashi Nakamura$^{1}$, Kazumi Kashiyama$^{2}$, Daisuke Nakauchi$^{1}$, Yudai Suwa$^{3}$, \\ Takanori Sakamoto$^{4}$, and Nobuyuki Kawai$^{5}$}

\altaffiltext{1}{Department of Physics, Kyoto University, Oiwake-cho, Kitashirakawa, Sakyo-ku,Kyoto 606-8502, Japan} 
\altaffiltext{2}{Department of Astronomy \& Astrophysics; Department of Physics; Center for Particle \& Gravitational Astrophysics; Pennsylvania State University, University Park, PA 16802, USA}
\altaffiltext{3}{Yukawa Institute for Theoretical Physics, Kyoto University, Oiwake-cho, Kitashirakawa, Sakyo-ku, Kyoto 606-8502, Japan}
\altaffiltext{4}{Department of Physics and  Mathematics, College of Science and Engineering, Aoyama Gakuin University, 5-10-1 Fuchinobe, Chuo-ku, Sagamihara-shi, Kanagawa 252-5258, Japan}
\altaffiltext{5}{Department of Physics, Tokyo Insititute of Technology, 2-12-1 Ookayama, Meguro-ku, Tokyo 152-8551, Japan}

\begin{abstract}
We investigate the possible origin of extended emissions (EEs) of short gamma-ray bursts 
with an isotropic energy of $\sim 10^{50\mbox{-}51} \ \rm erg$ and a duration of a few 10 s to $\sim 100 \ \rm s$, 
based on a compact binary (neutron star (NS)-NS or NS-black hole (BH)) merger scenario. 
We analyze the evolution of magnetized neutrino-dominated accretion disks of mass $\sim 0.1 \ M_\odot$ around BHs formed after the mergers,
and estimate the power of relativistic outflows via the Blandford-Znajek (BZ) process.   
We show that a rotation energy of the BH up to $\gtrsim 10^{52} \ \rm erg$ can be extracted with an observed time scale of $\gtrsim 30 (1+z)\ \rm s$ with a relatively small disk viscosity parameter of $\alpha < 0.01$.   
Such a BZ power dissipates by clashing with non-relativistic pre-ejected matter of mass $M \sim 10^{-(2\mbox{-}4)} \ M_\odot$, and forms a mildly relativistic fireball.  
We show that the dissipative photospheric emissions from such fireballs are likely in the soft X-ray band ($1\mbox{-}10 \ \rm keV$) for $M \sim 10^{-2} M_\odot$ possibly in NS-NS mergers, 
and in the BAT band ($15\mbox{-}150 \ \rm keV$) for $M \sim 10^{-4} M_\odot$ possibly in NS-BH mergers.  
In the former case, such soft EEs can provide a good chance of $\sim  {\rm 6 \ yr^{-1}} \ (\Delta \Omega_{softEE}/4\pi)  \ ({\cal R}_{GW}/40 \ {\rm yr^{-1}})$ 
for simultaneous detections of the gravitational waves with a $\sim 0.1^\circ$ angular resolution by soft X-ray survey facilities like Wide-Field MAXI.  
Here, $\Delta \Omega_{softEE}$ is the beaming factor of the soft EEs and ${\cal R}_{GW}$ is the NS-NS merger rate detectable by advanced LIGO, advanced Virgo, and KAGRA.  
\end{abstract}

 \maketitle

\section{Introduction}
Short gamma-ray bursts (SGRBs) are usually defined by the prompt duration, i.e., gamma-ray bursts (GRBs) with $T_{90} < 2 \ \rm s$ \citep{Kouveliotou1993}. 
A significant fraction of the SGRBs is accompanied by longer duration (up to $\sim 100 \ \rm s$) extended emissions (EEs)~\citep{Norris_Bonnell_2006}. 
The isotropic energy of the prompt spike is $E_{iso, PS} \sim 10^{50\mbox{-}51} \ \rm erg$~\citep{Tsutsui2013}, 
while that of EE is $E_{iso, EE} \sim 10^{50\mbox{-}51} \ \rm erg$ \citep{Sakamoto2011}. 
Interestingly, in some cases, observed fluences of the EEs are even larger than those of the prompt spikes, e.g., $E_{iso, EE} \sim 2.7 \times E_{iso, PS}$ for SGRB 050709 \citep{Villasenor2005} 
and $E_{iso, EE} \sim 30 \times E_{iso, PS}$ for SGRB 080503 \citep{Perley2009}. 
\cite{Fong2013} found that $\sim 25\%$ of {\it Swift} BAT SGRBs have EEs in the X-ray band (see their Table. 3). 
On the other hand, \cite{Bostanc2013} searched EEs in 296 BATSE SGRBs and found that  the fraction of EEs is $\sim 7\%$, 
where they pointed out that this fraction should be regarded as the minimum value since dim and/or softer EEs cannot be detected by BATSE. 
In fact, BATSE measures fluence above $20 \ \rm keV$ while BAT does down to $15 \ \rm keV$, 
and the 5 keV lower threshold energy yields an 18\% increase in the EE population. 
This suggests that a decrease in the threshold energy, say, as low as $\sim 1 \ \rm keV$, might yield a dramatic increase in the EE population.

Here, we consider the possible origin of such EEs based on the compact binary (neutron star (NS)-NS or NS-black hole (BH)) merger scenario 
\citep{Paczynski1986, Goodman1986, Eichler1989, Narayan1992}. 
If the maximum mass of a non-rotating NS is smaller than $\sim 2.5M_\odot$, the final outcome of such a merger will be a Kerr BH with a mass of $M_{BH}\sim 3 \ M_\odot$ 
and a spin parameter of $q=a/M_{BH} <  0.8$ with an accretion disk with a mass of $\sim 0.1 \ M_\odot$ and possibly beyond, and a neutron-rich ejecta with a mass of $M \sim 10^{-(2\mbox{-}4)} \ M_\odot$ 
with an expanding velocity of $v_{exp} \sim 0.1c$ \citep{Shibata2006, Kiuchi2009, Rezzolla2010, Hotokezaka2011,Hotokezaka2013}. 
Here, we adopt this situation. 
If the maximum mass of a non-rotating NS is larger than $\sim 2.5M_\odot$,  
a rapidly rotating massive NS will be the final outcome and the magnetar activities may be responsible for the prompt spike and EEs of the SGRBs as well as other electromagnetic counterparts
\citep{Usov1992, Zhang2001, Gao2006, Metzger2008, Bucciantini2012, Gompertz2013, Zhang2013}. 

In this paper, we consider the huge rotational energy of a Kerr BH to be the intrinsic energy budget of the EE~\citep[see also][]{Fan2005,Rosswog2007,Lee2009,Barkov2011}. 
The mass formula of the Kerr BH with a gravitational mass of $M_{BH}$ and an angular momentum $J$ is written as \citep{MTW1973}
\begin{equation}
M_{BH}^2=M_{ir}^2+\frac{J^2}{4M_{ir}^2},
\end{equation}
where $M_{ir}$ is the irreducible mass of the Kerr BH. 
Writing $J = aM_{BH} = qM_{BH}^2$, we have
\begin{equation}\label{eq:mass}
M_{BH}^2=M_{ir}^2+\frac{M^4q^2}{4M_{ir}^2},
\end{equation}
where $a$ is the well known Kerr parameter and $q = a/M_{BH} < 1$.
Eq. (\ref{eq:mass}) is rewritten as 
\begin{equation}
M_{BH}^2=\frac{2M_{ir}^2}{1+\sqrt{1-q^2}}.
\end{equation}
Then, the energy available by the extraction of the angular momentum of the Kerr BH is given by
\begin{equation}
\Delta E = M_{BH} \left( 1-\sqrt{\frac{1+\sqrt{1-q^2}}{2}} \right).
\end{equation}
For $q=0.5$, for example, $\Delta E$ is given by
\begin{equation}\label{eq:del_e}
\Delta E=1.84\times 10^{53} \ {\rm erg} \ \left( \frac{M_{BH}}{3 \ M_\odot} \right).
\end{equation}
Therefore, for an $\sim 100\%$ radiation efficiency, only $\sim 1\%$ of the rotational energy of the BH enables fueling of the EE even if the emission is isotropic. 
The problem is how to extract the rotational energy of the Kerr BH on a timescale of a few 10 s to $\sim 100 \ \rm s$.

One of the plausible mechanisms for extracting the rotation energy of the Kerr BH is the Blandford-Znajek (BZ) process \citep{Blandford1977}. 
From the results of numerical simulations, \cite{Penna2013} found that within factors of order unity, the BZ power is expressed in units of $c=G=1$ by
\begin{equation}
L^{BZ}=\frac{1}{6\pi}\Omega_H^2\Phi^2,
\end{equation}
where 
\begin{equation}
\Omega_H= M_{BH}^{-1} \times \frac{q}{(1+\sqrt{1-q^2})^2+q^2}
\end{equation}
and 
\begin{equation}
\Phi=\pi M_{BH}^2(1+\sqrt{1-q^2})^2B 
\end{equation}
is the magnetic flux threading the horizon with $B$ being the strength of the magnetic field formed by the disk around the Kerr BH.
Recovering $c$ and $G$, we have
\begin{equation}
L^{BZ}=\frac{\pi}{6}\left[\frac{q(1+\sqrt{1-q^2})^2}{(1+\sqrt{1-q^2})^2+q^2}\right]^2 \left(\frac{GM_{BH}}{c^2} \right)^2 c B^2.
\end{equation}
For $q=0.5$, for example, $L^{BZ}$ is expressed as
\begin{equation}\label{eq:l_bz}
L^{BZ} = 6.6\times 10^{50} \ {\rm erg \ s^{-1}} \ \left( \frac{M_{BH}}{3 \ M_\odot} \right)^2 \left(\frac{B}{10^{15} \ {\rm G}} \right)^2.
\end{equation}
Dividing Eq. (\ref{eq:del_e}) by Eq. (\ref{eq:l_bz}), we have the characteristic time $\delta t_{BZ}$ as
\begin{equation}
\delta t_{BZ} = 2.8\times 10^2 \ {\rm s} \ \left( \frac{M_{BH}}{3 \ M_\odot} \right)^{-1} \left( \frac{B}{10^{15}\ {\rm G}} \right)^{-2}.
\end{equation}
This shows that if an accretion disk with $B \sim 10^{15} \ \rm G$ and an accretion time $\sim 100 \ \rm s$ exists, up to $\sim 10^{53}$ erg can be extracted from the Kerr BH. 
This is just the timescale of the EEs and only $\sim 1\%$ efficiency is enough to explain the emissions even if they are isotropic. 

This paper is organized as follows. In \S 2, we analyze the time evolution of the neutrino-dominated accretion disk with finite mass and angular momentum around the BH,  
and estimate the resultant BZ power and the duration.  
In \S 3, we consider the interaction of the BZ jets with the pre-ejected matter ($M \sim 10^{-(2\mbox{-}4)} \ M_\odot$) with an expanding velocity of $v_{exp} \sim 0.1c$, 
which produces mildly relativistic fireballs. 
We calculate the dissipative photospheric emissions from such fireballs. 
There, we also argue the detectability and the association with the observed EEs. 
\S 4 is devoted to the discussion. 
We use $Q_x = Q/10^x$ in CGS units unless otherwise noted.

\section{Blandford-Znajek jets from black-hole tori formed after compact binary mergers}
 
\cite{Chen2007} calculated the steady-state solutions of a neutrino-dominated accretion disk around a Kerr BH, 
which is also the case in our setup at the initial stage.  
They assumed an accretion rate of $\dot{M(t)} = $ constant, but took into account the full neutrino process and the Kerr geometry.
Their important conclusions are that (i) the pressure is dominated by baryons with $p = (\rho/m_p)kT$,
(ii) the disk is neutron dominated so that the electron fraction is as small as $Y_e \sim 0.1$, 
(iii) the degeneracy of the electron is at most mild because, if the degeneracy is high, the neutrino cooling is lowered to increase the temperature, 
(iv) there is an ignition accretion rate for the neutrino cooling disk that is proportional to $\alpha^{5/3}$ where $\alpha$ is the parameter in the so-called  $\alpha$-disk model \citep{Shakura1973}.  
\cite{Kawanaka2013} performed simpler Newtonian calculations of such disks both numerically and analytically. 
One of the  important conclusions is that  their analytical model fits well with the numerical ones by \cite{Chen2007}.
Therefore, here we adopt a simple Newtonian analytical model  to mimic the neutrino-dominated accretion disk around the Kerr BH.
One of the big differences from \cite{Kawanaka2013} is that we take into account the time variation of the accretion rate for the finite disk mass and 
the finite disk angular momentum while they considered a constant accretion rate.
 
The structure of the accretion disk can be derived from \citep{Kawanaka2013} 
\begin{eqnarray}
\dot{M}&=&-2\pi r \times 2\rho hv_r,\\
2\alpha hp&=&\frac{\dot{M}\Omega}{2\pi},\\
p&=&\rho\Omega^2h^2,
\end{eqnarray}
where $\dot{M}, r, \rho,  h, v_r, \alpha,  p$ and  $ \Omega$ are the accretion rate,  $r$ in cylindrical coordinates, the density, the half 
thickness of the disk, the infalling velocity, the $\alpha$ parameter, the pressure, and the angular frequency of the disk, respectively.
We can express $h, p, v_r$ by $\rho$ as
\begin{eqnarray}
h &=& \left(\frac{\dot{M}}{4\pi\alpha\Omega\rho}\right)^{1/3} \nonumber  \\
&=& 7.02\times10^8 \ {\rm cm} \ \alpha_{-1}^{-1/3} \rho_0^{-1/3}, \notag \\
&& \times \left(\frac{\dot{M}}{10^{-3}M_\odot s^{-1}}\right)^{1/3}\left(\frac{M_{BH}}{3 \ M_\odot}\right)^{1/3}\left(\frac{r}{r_{ISCO}}\right)^{1/2},
\end{eqnarray}
\begin{eqnarray}
p &=& \Omega^{4/3}\rho^{1/3}\left(\frac{\dot{M}}{4\pi\alpha}\right)^{2/3}\nonumber \\
&=&1.04\times10^{25} \ {\rm dyn \ cm^{-2}} \ \alpha_{-1}^{-2/3} \rho_0^{1/3}, \notag \\
&& \times \left(\frac{\dot{M}}{10^{-3}M_\odot s^{-1}}\right)^{2/3}\left(\frac{M_{BH}}{3 \ M_\odot}\right)^{-4/3}\left(\frac{r}{r_{ISCO}}\right)^{-2}, \label{eq:p_gene}
\end{eqnarray}
\begin{eqnarray}
v_r &=& -\left(\frac{\dot{M}}{4\pi\rho}\right)^{2/3}\left(\alpha\Omega\right)^{1/3}\frac{1}{r} \nonumber \\
&=& -8.49\times10^{13} \ {\rm cm \ s^{-1}} \ \alpha_{-1}^{1/3} \rho_0^{-2/3}, \notag \\
&& \times \left(\frac{\dot{M}}{10^{-3}M_\odot s^{-1}}\right)^{2/3}\left(\frac{M_{BH}}{3 \ M_\odot}\right)^{-4/3}\left(\frac{r}{r_{ISCO}}\right)^{-3/2},
\end{eqnarray}
where  
\begin{equation}
r_{ISCO} = 6GM_{BH}/c^2,
\end{equation}
is the innermost stable circular orbit (ISCO).\footnote{In fact, the location of the ISCO depends on $q=a/M$. 
However we are using Newtonian gravity so that the exact treatment of ISCO is not possible. 
One can take into account the change of ISCO by putting different value of $r/6GM_{BH}/c^2$ in all the equations. 
In this case, various quantities are modified by powers of the above factor, but our results do not change qualitatively.}
The density $\rho$ can be determined by the energy equation and the equation of state.
Denoting the cooling rate $\dot{q}$, the energy balance is expressed as
\begin{eqnarray}
\dot{q}&=& \frac{3GM_{BH}\dot{M}}{4\pi r^3 \times 2h}\\
&=&7.16\times10^{27} \ {\rm erg \ cm^{-3} \ s^{-1}} \ \alpha_{-1}^{1/3} \rho_0^{1/3} \notag \\
&& \times \left(\frac{\dot{M}}{10^{-3}M_\odot s^{-1}}\right)^{2/3}\left(\frac{M_{BH}}{3 \ M_\odot}\right)^{-7/3}\left(\frac{r}{r_{ISCO}}\right)^{-7/2}. \label{eq:q_eq}
\end{eqnarray}

As for the energy loss rate, we consider two neutrino cooling processes relevant to the accretion disk we are interested in as \citep{Itoh1989, Popham1999}
\begin{eqnarray}
\dot{q}_{\rm  URCA} &=& 9\times 10^{-43} \  {\rm erg \ cm^{-3} \ s^{-1}} \ \rho_0 T_0^6, \label{eq:q_urca}\\
\dot{q}_{\rm  pair} &=& 5\times 10^{-66} \  {\rm erg \ cm^{-3} \ s^{-1}} \  T_0^9,
\end{eqnarray}
where $T$ is the temperature in units of [K]. 
The URCA process is $p+e^-\rightarrow n+\nu_e, n+e^+\rightarrow p+\bar{\nu}_e$ and the pair neutrino process is $e^-+e^+\rightarrow \nu_e+\bar{\nu}_e$. 
The URCA process dominates over the pair process for $\rho_0 >5.5\times 10^{-24}T_0^3$.
As for the pressure of the matter we should consider 
\begin{equation}
p = \frac{\rho kT}{m_p}  \ \ \  {\rm (gas ~pressure)},  \label{eq:eos}
\end{equation}
\begin{equation}
p = \frac{1}{3}a_{rad} T^4  \ \ \  {\rm (pressure~ by~ radiation)},
\end{equation}
\begin{equation}
p = \frac{2\pi ch}{3}\left(\frac{3Y_e\rho}{8\pi m_p}\right)^{4/3} \ \ \  {\rm (degenerate~electron)}, 
\end{equation}
where $a_{rad} = 7.57 \times 10^{-15} \ \rm erg \ cm^{-3} \ K^{-4}$ is the radiation constant.
The relativistically degenerate pressure dominates over the gas and the radiation pressure for $\rho >2.3\times 10^{-22} \ {\rm g \ cm^{-3}} \ (Y_e/0.5)^{-4} T_0^3$ 
while the gas pressure dominates over the radiation pressure for $\rho >3.06\times 10^{-23} \ {\rm g \ cm^{-3}} \ T_0^3$ 

Let us consider the case for  $3.06\times 10^{-23} \ T_0^3 < \rho_0 < 1.4\times  10^{-19} \ T_0^3$, where the pressure is determined by gas pressure  
and the cooling process is dominated by the URCA  process so that Eqs. (\ref{eq:q_eq}) and (\ref{eq:q_urca}) give 
\begin{eqnarray}
&& 7.18\times10^{27} \ \alpha_{-1}^{1/3} \rho_0^{1/3} \notag \\
&& \times  \left(\frac{\dot{M}}{10^{-3}M_\odot s^{-1}}\right)^{2/3}\left(\frac{M_{BH}}{3 \ M_\odot}\right)^{-7/3}\left(\frac{r}{r_{ISCO}}\right)^{-7/2} \label{eq:cool_eq} \notag \\
&=& 9\times 10^{-43} \ \rho_0 T_0^6, 
\end{eqnarray}
with $p = \rho kT/m_p$. 

From Eqs. (\ref{eq:p_gene}), (\ref{eq:eos}), and (\ref{eq:cool_eq}), $\rho$ is expressed as
\begin{eqnarray}
\rho &=& 6.47\times 10^{9} \ {\rm g \ cm^{-3}} \ \alpha_{-1}^{-13/10} \notag \\
&& \times \left(\frac{\dot{M}}{10^{-3}M_\odot s^{-1}}\right)\left(\frac{M_{BH}}{3 \ M_\odot}\right)^{-17/10}\left(\frac{r}{r_{ISCO}}\right)^{-51/20}.
\end{eqnarray}
We then have 
\begin{eqnarray}
h &=& 3.77\times 10^{5} \ {\rm  cm} \ \alpha_{-1}^{1/10} \notag \\
&& \times \left(\frac{M_{BH}}{3 \ M_\odot}\right)^{9/10}\left(\frac{r}{r_{ISCO}}\right)^{27/20},
\end{eqnarray}
\begin{eqnarray}
p &=& 1.94\times 10^{28} \ {\rm  dyn \ cm^{-2}} \ \alpha_{-1}^{-11/10} \notag \\
&& \times \left(\frac{\dot{M}}{10^{-3}M_\odot s^{-1}}\right)\left(\frac{M_{BH}}{3 \ M_\odot}\right)^{-19/10}\left(\frac{r}{r_{ISCO}}\right)^{-57/20}, \label{eq:p}
\end{eqnarray}
\begin{eqnarray}
T &=& 3.63\times 10^{10} \ {\rm K} \ \alpha_{-1}^{1/5}\left(\frac{M_{BH}}{3 \ M_\odot}\right)^{-1/5}\left(\frac{r}{r_{ISCO}}\right)^{-3/10},
\end{eqnarray}
\begin{equation}
v_r = -2.45\times 10^{7} \ {\rm cm \ s^{-1}} \ \alpha_{-1}^{6/5}\left(\frac{M_{BH}}{3 \ M_\odot}\right)^{-1/5}\left(\frac{r}{r_{ISCO}}\right)^{1/5}, \label{eq:v_r}
\end{equation}
\begin{eqnarray}
v_r/v_{kep} &=& -2.00\times 10^{-3} \ \alpha_{-1}^{6/5}, \notag \\
&& \times \left(\frac{M_{BH}}{3 \ M_\odot}\right)^{-1/5}\left(\frac{r}{r_{ISCO}}\right)^{7/10}, \label{eq:v_ratio}
\end{eqnarray}
\begin{eqnarray}
v_{kep} &=& 1.22\times 10^{10} \ {\rm cm \ s^{-1}} \ \left(\frac{r}{r_{ISCO}}\right)^{-1/2}.
\end{eqnarray}
Then, the surface density of the disk ($\Sigma$) is given by
\begin{eqnarray}
\Sigma = 2\rho h &=& 4.88\times 10^{15} \ {\rm g \ cm^{-2}} \ \alpha_{-1}^{-6/5} \notag \\
&& \times \left(\frac{\dot{M}}{10^{-3}M_\odot s^{-1}}\right)\left(\frac{M_{BH}}{3 \ M_\odot}\right)^{-4/5}\left(\frac{r}{r_{ISCO}}\right)^{-6/5}.
\end{eqnarray}

Let us introduce the coordinate $x$ by $r = x \times r_{ISCO}$. 
We assume here that the disk has a minimum and a maximum $x$ as $x_{min}=1$ and $x_{max}$, respectively. 
Then for a given total mass $m_d$ and the total angular momentum $J_t$, we have
\begin{eqnarray}
m_d &=& 1.36\times 10^{-4} \ M_\odot \alpha_{-1}^{-6/5} (x_{max}^{4/5}-1) \notag \\
&& \times \left(\frac{\dot{M}}{10^{-3}M_\odot s^{-1}}\right)\left(\frac{M_{BH}}{3 \ M_\odot}\right)^{6/5}, \label{eq:m_d}\\
J_t &=& 5.48\times10^{45} \ {\rm g \ cm^2 \ s^{-1}} \ \alpha_{-1}^{-6/5} (x_{max}^{13/10}-1) \notag \\
&& \times \left(\frac{\dot{M}}{10^{-3}M_\odot s^{-1}}\right)\left(\frac{M_{BH}}{3 \ M_\odot}\right)^{11/5}. \label{eq:j_t}
\end{eqnarray}
For a given value of $m_d$, $J_t$, $\alpha$, and $M_{BH}$, from Eqs. (\ref{eq:m_d}) and (\ref{eq:j_t}), we can determine 
$\dot{M}$ and $x_{max}$ in general.  
Let us define a new variable $\beta$ by  $J_t = m_d\beta \sqrt{GM_{BH} r_{ISCO}}$, 
that is, the mean value of the angular momentum is $\beta( > 1)$ times  the minimum value of the specific angular momentum at the ISCO. 
Then, $x_{max}$ is determined by 
\begin{equation}\label{eq:x_max}
\frac{x_{max}^{13/10}-1}{x_{max}^{4/5}-1}=1.625\beta.
\end{equation}
It is easily shown that the left-hand side of Eq. (\ref{eq:x_max}) is a monotonically increasing function for $x_{max} > 1$ and has a minimum value $1.625$ at  $x_{max} =1$ 
so that for an arbitrary value of $\beta > 1$, there is an unique solution $x_{max} > 1$.
In the accretion process, the total angular momentum of the system  should be conserved in our case, since the Kerr metric has a rotational Killing vector, i.e., a stationary axisymmetric system. 
Some of the angular momentum is absorbed by the BH from the ISCO so that $\beta$ increases as a function of time. 
Note that the spin up of the BH due to accretion is negligible in our case. 
If we denote the mass and the angular momentum of the accreted blob into the BH as $\Delta m_d( < 0)$ and $\Delta J=\Delta m_d\sqrt{GM_{BH} r_{ISCO}}$, we have
\begin{equation}\label{eq:beta}
\Delta \beta=(1-\beta)\frac{\Delta m_d}{m_d} >0.
\end{equation}
The solution of Eq. (\ref{eq:beta}) is given by
\begin{equation}\label{eq:beta_int}
\beta=1+(\beta_0-1)\frac{m_d^0}{m_d},
\end{equation}
where $\beta_0$ and $m_d^0$ are the initial values of $\beta$ and $m_d$, respectively. 
Therefore, $\beta\gg 1$ in the later phase of the accretion so that $x_{max}\gg 1$ and the $x_{max} \approx (1.6\beta)^2 \approx [1.6(\beta_0-1) \times m_d^0/m_d]^2$ will be a good approximation. 
Inserting this expression to Eq. (\ref{eq:m_d}), we have
\begin{equation}\label{eq:d_m_d}
\frac{\dot{m_d}}{m_d^{13/5}} = -3.47 \ {\rm s^{-1}} \ [m_d^0(\beta_0-1)]^{-8/5}\alpha_{-1}^{6/5}\left(\frac{M_{BH}}{3 \ M_\odot}\right)^{-6/5}.
\end{equation}
Integration of Eq. (\ref{eq:d_m_d}) yields
\begin{eqnarray}
m_d&=&\frac{m_d^0}{(t/A+1)^{5/8}}, \\
A&=&0.18 \ {\rm s} \ (\beta_0-1)^{8/5} \alpha_{-1}^{-6/5} \left(\frac{M_{BH}}{3 \ M_\odot}\right)^{6/5}, \label{eq:A}\\
\dot{m}_d&=&-\frac{5m_d^0}{8A(t/A+1)^{13/8}} \label{eq:m_d_ana}.
\end{eqnarray}

The method we adopted here to solve the evolution of the accretion disk is similar to the quasi-static evolution of the star where the nuclear timescale is much longer than the free fall time 
so that at each time, the star can be regarded  in gravitational equilibrium \citep[see, e.g.,][]{Kippenhahn1994}. 
In our case, from Eq. (\ref{eq:v_ratio}), the accretion velocity is much smaller than the Kepler velocity which determines the dynamical timescale so that we can regard the disk as stationary at each time.  
The decrease in total mass and angular momentum can be regarded as a decrease in total nuclear energy and a change of the composition in the stellar evolution case, 
which are very slowly changing in a dynamical timescale. 

\begin{figure}
\includegraphics[width = 150mm]{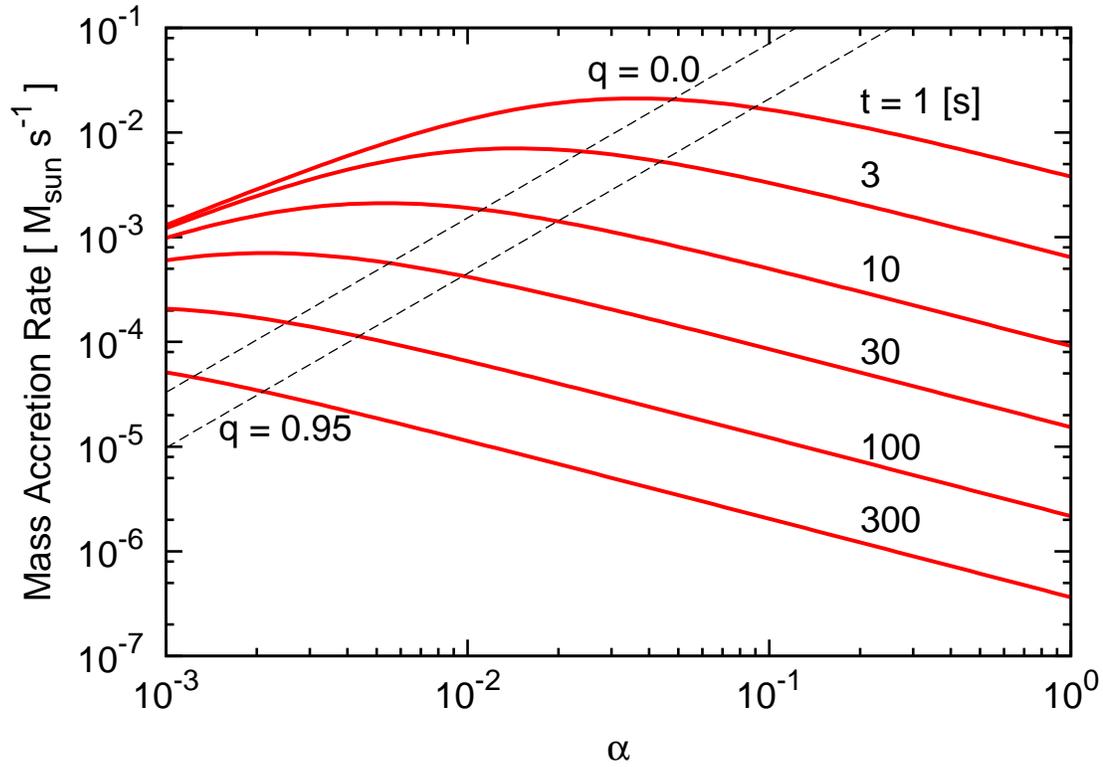}
\caption{Accretion rate as a function of $\alpha$ for typical time such as $1 \ \rm s$, $3 \ \rm s$, $10 \ \rm s$, $30 \ \rm s$, $100 \ \rm s$ and $300 \ \rm s$ 
for $m_d^0 = 0.1 \ M_\odot$, $\beta_0 = 2$, $M_{BH}= 3 \ M_\odot$. 
The dashed lines are neutrino cooling ignition  accretion rate obtained by \cite{Chen2007} for $q=0$ and $q=0.95$ (Eq. 42 of their paper). 
Neutrino cooling is  effective only above these dashed lines.}
\label{Mdot-1}
\end{figure}

\begin{figure}
\includegraphics[width = 150mm]{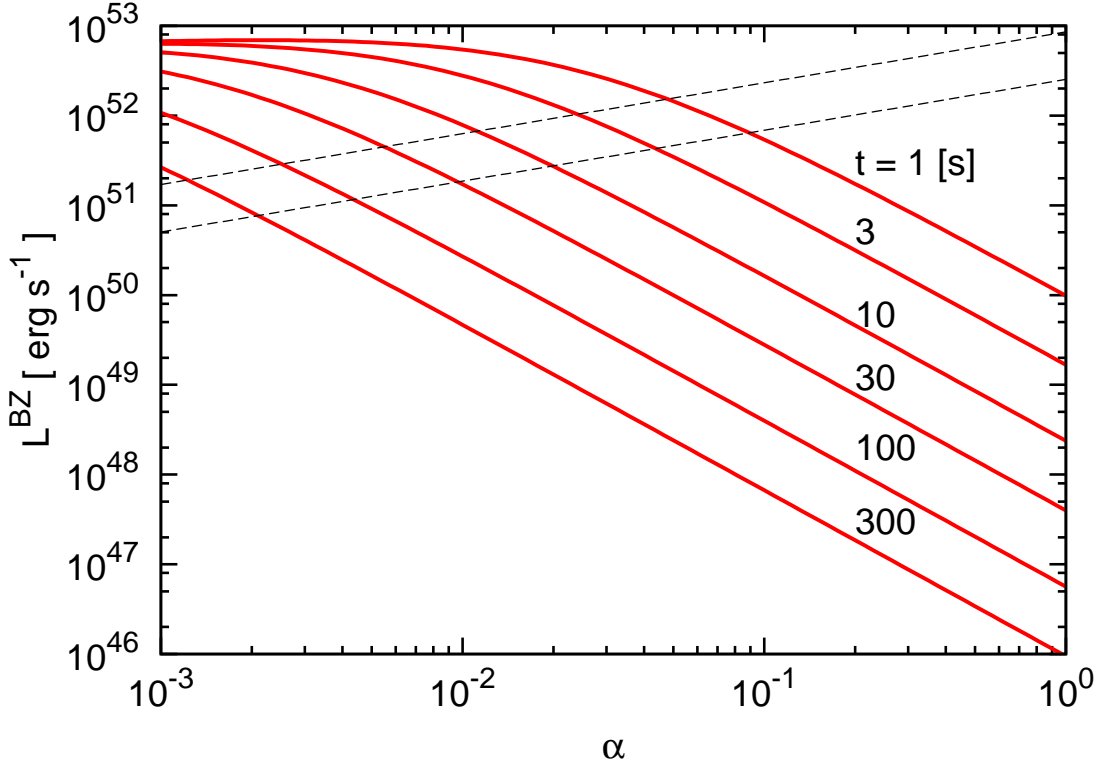}
\caption{The Blandford-Znajek (BZ) luminosity as a function of $\alpha$ for typical time such as $1 \ \rm s$, $3 \ \rm s$, $10 \ \rm s$, $30 \ \rm s$, $100 \ \rm s$ and $300 \ \rm s$ 
for $m_d^0=0.1 \ M_\odot$, $\beta_0=2$, $M_{BH}=3 \ M_\odot$, and $q = 0.5$. 
The magnetic field at the horizon is assumed to be determined by $B^2=8\pi p_{ISCO}$ where $p_{ISCO} $ is the pressure of the disk at the ISCO ($r=6GM_{BH}/c^2$) in Eq. (\ref{eq:p}). 
The BZ luminosity is then given by Eq. (\ref{eq:l_bz}).
The dashed lines show the neutrino cooling ignition accretion rates for $q=0$ (upper) and $q=0.95$ (lower). 
} 
\label{L_bz}
\end{figure}

Let us assume that $m_d^0 = 0.1 \ M_\odot$, $\beta_0 = 2$, $M_{BH} = 3 \ M_\odot$, and $q = 0.5$ as suggested by numerical relativity calculations
\citep{Shibata2006, Kiuchi2009, Rezzolla2010, Hotokezaka2011,Hotokezaka2013}.  
Solid lines in Fig. \ref{Mdot-1} show  the accretion rates as a function of $\alpha$ for the representative time such as 1 s, 3 s, 10 s, 30 s, 100 s, and 300 s. 
We are interested in the late time behavior ($t > 30$ s) where the accretion rate decreases as a function of  $\alpha$ for a fixed time. 
This is because $A$ in Eq. (\ref{eq:A}) is smaller for larger $\alpha$ so that the accretion rate is smaller. 
Physically, the accreting velocity is larger for larger viscosity as is clear from Eq. (\ref{eq:v_r}), where the consumption of the disk mass is faster.  
The dashed lines are the neutrino cooling ignition  accretion rate obtained by \cite{Chen2007} for $q = 0$ and $q = 0.95$ (Eq. 42 of their paper). 
Neutrino cooling is  effective only above these dashed lines. 
In Fig. \ref{L_bz}, we show BZ luminosity as a function of $\alpha$ for typical time such as 1 s, 3 s, 10 s, 30 s, 100 s, and 300 s 
for $m_d^0 = 0.1 \ M_\odot$, $\beta_0 = 2$, $M_{BH} = 3 \ M_\odot$, and $q = 0.5$. 
The magnetic field at the horizon is assumed to be determined by 
\begin{equation}
\frac{B^2}{8\pi} = p_{\rm ISCO},
\end{equation}
where $ p_{ISCO}$ is the pressure of the disk at the ISCO in Eq. (\ref{eq:p}). 
The BZ luminosity is then given by Eq. (\ref{eq:l_bz}). 
We clearly see that the BZ luminosity is higher for smaller $\alpha$ for the same time. 
This comes from both the strong $\alpha$ dependence of the $p_{ISCO}$ in Eq. (\ref{eq:p}) and the accretion rate in Fig. \ref{Mdot-1}. 
Physically, if the viscosity is low, the accretion rate decreases slowly and the pressure is high due to the accumulation of matter, which yields a strong magnetic field in our scenario. 
All these effects result in a larger BZ luminosity for a smaller $\alpha$.
  
\begin{figure}
\includegraphics[width = 150mm]{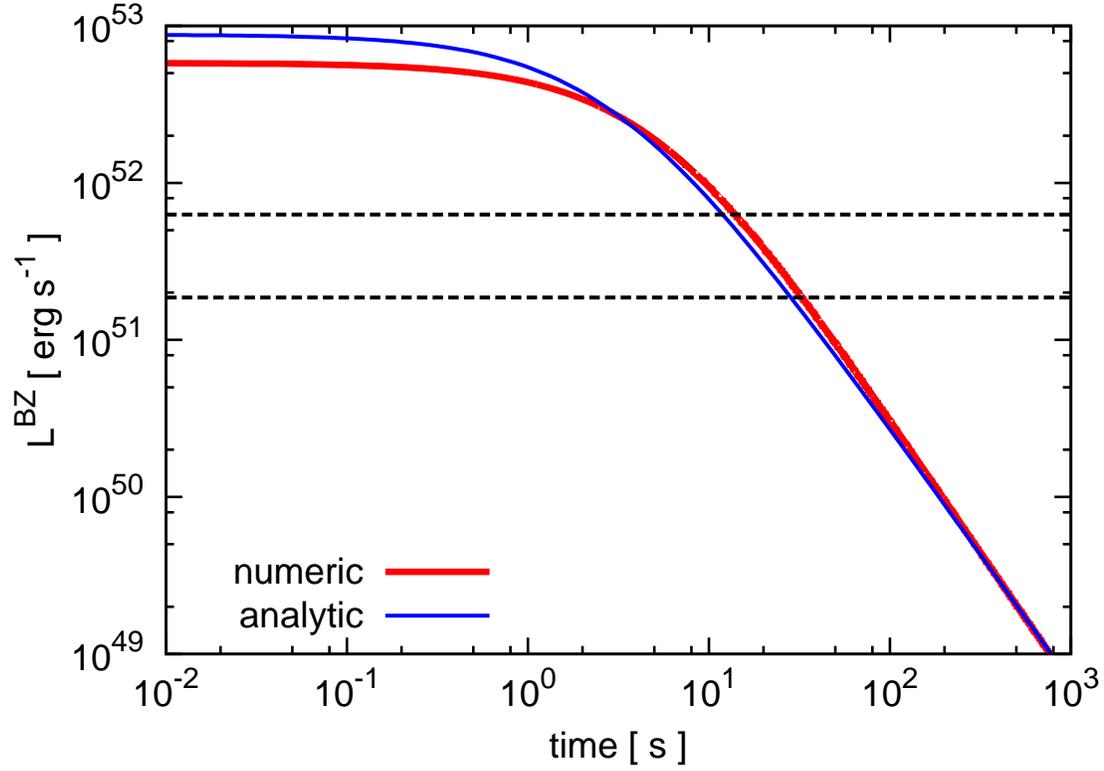}
\caption{Time evolution of the luminosity for $m_d^0 = 0.1 \ M_\odot$, $\beta_0 = 2$, $M_{BH} = 3 \ M_\odot$, $q = 0.5$, and $\alpha=0.01$. 
The red and blue solid lines show the numerical exact solution and the analytic approximation, respectively. 
The dashed lines show the neutrino cooling ignition accretion rates for $q=0$ (upper) and $q=0.95$ (lower). 
We see that the analytic approximation only  overestimates the luminosity slightly so that  we can justify the use of the analytic approximate solution to Eq. (\ref{eq:x_max}).} 
\label{check_LC}
\end{figure}

One might suspect that our approximation ($x_{max} \gg 1$) to the solution of Eq. (\ref{eq:x_max}) for given $\beta$ affects the result. 
To check this, we show in Fig. \ref{check_LC} the time evolution of the luminosity for $m_d^0 = 0.1 \ M_\odot$, $\beta_0 = 2$, $M_{BH} = 3 \ M_\odot$, $q = 0.5$, and $\alpha=0.01$.
The red and blue solid lines show the numerical solution, which is obtained by integrating Eqs. (\ref{eq:m_d}), (\ref{eq:j_t}), and (\ref{eq:x_max}) directly over time, the analytic approximation (Eq. \ref{eq:m_d_ana}). 
The dashed horizontal lines show the neutrino cooling ignition accretion rates for $q=0$ (upper) and $q=0.95$ (lower), respectively. 
We see that the analytic approximation only slightly overestimates the luminosity. 
We can justify the use of the analytic approximate solution to Eq. (\ref{eq:x_max}).
   
It is suggested that the typical $\alpha \sim 0.01\mbox{-}0.02$ from various numerical simulations~\citep{Davis2010,Guan2011,Blaes2011,Parkin2013,Jiang2013}.  
To have possible  isotropic EE, the typical luminosity of $\sim 10^{49} \ {\rm erg \ s^{-1}}$ is needed at $t \sim$ a few 10 s. 
From Fig. \ref{Mdot-1}, if $\alpha \lesssim 0.01$, which is relatively smaller than that suggested by various numerical simulations, 
the accretion rate is above the ignition rate for neutrino cooling up to the observed time of  $t \sim 30(1+z)$ s, and the luminosity can be above $\sim 10^{50}\ {\rm erg \ s^{-1}}$ 
so that enough luminosity for EEs seems to be obtained via the BZ mechanism.
\if
However, from Fig. \ref{Mdot-1}, the accretion rate is above the ignition rate for neutrino cooling only up to $t \sim 30$ s. 
In this case the duration of EE is at most $30$ s. 
To increase the duration of EE, we need smaller $\alpha$ than 0.01. 
When can the value of $\alpha$ become smaller?  
If the origin of viscosity is magnetic turbulence from the magneto rotational instability, there is a big difference from the usual situations, that is, the matter is neutron rich ($Y_e \sim 0.1$). 
Since neutron does not directly couple to the magnetic field, and $Y_e \sim 0.1$, only 10\% of the mass feels the viscosity first. 
Note that, for $T\sim 10^{10} \ {\rm K}$, $B\sim 10^{15} \ {\rm G}$, $\rho \sim 10^{10} \ {\rm g \ cm^{-3}}$, 
which are typical values at the ISCO, the gyration radius of the proton $\sim 10^{-10}$ cm is comparable to the mean free path of p-n collision with the cross section of $\sim 1 \ {\rm barn}$. 
Then, the effective $\alpha_{eff}$ might be reduced, in principle, 10\% compared to the usual case of proton (= hydrogen) dominated gas, 
i.e., $\alpha \sim 0.01$ in the conventional definition could give $\alpha_{eff} \sim 0.001$. 
Then, from Fig. \ref{Mdot-1}, the duration can be even longer than $\sim 100$ s and the BZ luminosity can supply an enough energy for the isotropic EE.  
These arguments suggest that the rotational energy of the Kerr BHs formed in compact binary mergers might supply the energy sources of the EEs  
with the required duration even if the emissions are isotropic.
\fi
  
In Newtonian gravity, a numerical calculation of the accretion disk exists for the present problem, i.e., 
the time evolution of an accretion disk of mass $0.1 \ M_\odot$ after the merger of an NS-NS  binary \citep{Metzger2009}. 
Our  treatment also uses Newtonian gravity so that we can compare our analytic  results with their numerical calculations to confirm the quantitative agreement. 
In their calculations, the mass of the central BH is $3 \ M_\odot$ and they solved the time evolution of a z-direction integrated quantity such as the surface density $\Sigma$ with neutrino and advection cooling.  
The initial surface density is given as
\begin{equation}
\Sigma \propto (r/r_{d,0})^5\exp(-7r/r_{d,0}),
\end{equation}
with $r_{d,0} = 3\times 10^6 \ {\rm cm}\sim 6GM_{BH}/c^2$. 
Since $\Sigma r^2$ peaks at $r_{d,0}$, the initial specific angular momentum is $\sim \sqrt{GM_{BH} \times 6GM_{BH}/c^2}$. 
In our crude model, we assumed the disk boundary is at $r_{ISCO} = 6GM_{BH}/c^2$ while in their calculation inner edge (= disk boundary) is $10^6 \ {\rm cm} \sim 2GM_{BH}/c^2$ 
so that we define the coordinate $x$ by $r=x \times 2GM_{BH}/c^2$ in this paragraph. 
In the Newtonian calculation, there is no ISCO so that the minimum specific angular momentum of their calculation is $\sim \sqrt{GM_{BH} \times 2GM_{BH}/c^2}$ 
which is different from our model of  $\sqrt{GM_{BH}\times 6GM_{BH}/c^2}$. 
As for $\alpha$, they adopt 0.3 so that we need to rewrite Eqs. (\ref{eq:m_d}) and (\ref{eq:j_t}) as    
\begin{eqnarray}
m_d &=& 1.53\times 10^{-5} \ M_\odot (x_{max}^{4/5}-1) \notag \\
&& \times \left(\frac{\dot{M}}{10^{-3}M_\odot s^{-1}} \right)\left(\frac{M_{BH}}{3 \ M_\odot}\right)^{6/5},\\
J_t &=& 3.52\times10^{44} \ {\rm g \ cm^2 \ s^{-1}} \ (x_{max}^{13/10}-1) \notag \\
&& \times \left(\frac{\dot{M}}{10^{-3}M_\odot s^{-1}}\right)\left(\frac{M_{BH}}{3 \ M_\odot}\right)^{11/5}.
\end{eqnarray}
Defining $\beta$ by $J_t = m_d\beta \sqrt{GM_{BH} \times 2GM_{BH}/c^2}$, we have the same equation as Eq. (\ref{eq:x_max}). 
The argument for deriving equations corresponding to Eqs. (\ref{eq:beta}) and (\ref{eq:beta_int}) is also the same by changing $\Delta J = \Delta m_d\sqrt{GM_{BH} \times 2GM_{BH}/c^2}$ 
and $\beta_0\sim \sqrt{3}$, which gives  
\begin{equation}\label{eq:m_d_metzger}
\frac{\dot{m_d}}{m_d^{13/5}} = -51.0 \ {\rm s^{-1}} \ (m_d^0)^{-8/5}\left(\frac{M_{BH}}{3 \ M_\odot}\right)^{-6/5}.
\end{equation}
Integration of Eq. (\ref{eq:m_d_metzger}) yields
\begin{eqnarray}
m_d &=& 0.1\ M_\odot \left(\frac{t}{1.22\times10^{-2} \ {\rm s}}+1\right)^{-5/8},\\
\dot{m}_d&=&- 5.12 \ M_\odot \ {\rm s^{-1}} \  \left(\frac{t}{1.22\times10^{-2} \ {\rm s}} +1 \right)^{-13/8}. \label{eq:m_d_metzger_int}
\end{eqnarray} 

Now let us compare Eq. (\ref{eq:m_d_metzger_int}) with those of the numerical calculations by \cite{Metzger2009}. 
Their Fig. 3 shows an accretion rate at $r=10^6 \ {\rm cm}$ for $t = 0.01\ {\rm s}, 0.1 \ {\rm s}$ and $1\ {\rm s}$ are 
$1 \ M_\odot \ {\rm s^{-1}}$, $0.1 \ M_\odot \ {\rm s^{-1}}$ and $4.5\times 10^{-3} \ M_\odot \ {\rm s^{-1}}$, respectively. 
While Eq. (\ref{eq:m_d_metzger_int})  yields $1.93 \ M_\odot \ {\rm s^{-1}}$, $0.139 \ M_\odot \ {\rm s^{-1}}$ and $3.9\times 10^{-3} \ M_\odot \ {\rm s^{-1}}$. 
We can say that our analytic model agrees rather well with the numerical calculations at ISCO  especially for later times, which is indispensable for the use of our analytic model to study EEs. 
 
It has been shown that, in the late phase of the accretion of dense debris such as we consider here, 
the disk wind driven by energy injection via viscous heating and the recombination of nucleons into alpha-particles becomes relevant \citep{Metzger2008,Beloborodov2008,Metzger2009,Lee2009,Fernandez2013}. 
Although our calculation does not include this effect, since such outflows are predominantly triggered after the viscous timescale of the disk, our results can be still viable up to this timescale, 
e.g., $t \sim 30 (1+z)\ \rm s$ for $\alpha \lesssim 0.01$ (see Fig. 1). 
 
 \section{Extended X-ray emission as an electromagnetic counterpart of compact binary merger}
In the previous section, we showed that the rotational energy of the Kerr BH up to $E_{BZ} \sim 10^{52} \ \rm erg$ 
can be extracted as the Poynting outflow via the BZ process with a timescale of $\delta t_{BZ} \gtrsim 30 \ \rm s$ if the accretion of the debris $\sim 0.1 \ M_{\odot}$ occurs with $\alpha \lesssim 0.01$.
Hereafter, we argue the resultant emissions from such outflows and their detectability.  

In the course of compact binary mergers, a fraction of baryons of mass $M \sim 10^{-(2\mbox{-}4)} \ M_\odot$ can be ejected with an expansion velocity of $v_{exp} \gtrsim 0.1c$ \citep{Hotokezaka2013}. 
A certain duration after the merger, say $\delta t \sim 0.1\ \rm s$, the hypermassive NS collapses into a BH due to the loss of rotational support by emitting gravitational waves~(GWs) and/or Poynting fluxes. 
The Poynting outflow by the BZ process, which is relativistic, clashes with the pre-ejecta. 
The BZ outflow or jet will be more or less beamed and drill through the pre-ejecta, forming a hot plasma cocoon surrounding the jet.
Recently, such a  situation has been investigated numerically~\citep{Nagakura2014,Murguia2014}, 
although the jet injection timescale is set to be $< 1$ s, considering jets responsible for prompt emissions of SGRBs. 
These studies show that the jet dynamics are significantly affected by the pre-ejecta, 
especially for a jet luminosity of $\lesssim 10^{51} \ \rm erg \ s^{-1}$ and a pre-ejecta mass ejection rate of $\gtrsim 10^{-3} \  M_\odot \ s^{-1}$, 
which we are interested in here. 
In particular, if the jet launch is delayed more than $\delta t \gtrsim 0.1$ s from the pre-ejecta launch, 
such a jet will be choked in the ejecta and a significant fraction of its energy will be dissipated inside the ejecta, forming a cocoon fireball.
In our case, the assumed duration of the outflow injection is $\gtrsim 10$ s, thus, the BZ outflow would more easily penetrate the pre-ejecta
if the onset time of outflow injection is not delayed significantly. 
Nevertheless, even after penetration, a fraction of the BZ-outflow energy can dissipate at the interaction surface with the 
pre-ejecta or the cocoon, which typically occurs at 
\begin{equation}\label{eq:r_o}
r_o \sim (0.1\mbox{-}1) \times c \delta t_{BZ} \sim 3.0 \times 10^{11\mbox{-}12} \ {\rm cm}  \ \delta t_{BZ,2},   
\end{equation}
where $\delta t_{BZ}$ corresponds to the time $t$ in the previous section. 
Note that the BZ outflow is most likely magnetically dominated at the launching radius. 
In the case of magnetically dominated jets, the dynamics including the cocoon formation are different from those of hydrodynamical jets \citep[e.g.,][]{Bromberg2014}, 
and the energy dissipation process is rather uncertain.
Hereafter, we simply parameterize such a dissipation process by the fraction of dissipated energy, $\xi \leq 1$ and the beaming factor of the dissipation region, $\Delta \Omega$.

After dissipation, the heated pre-ejecta can be regarded as a fireball.  
The temperature can be estimated as $T'_o \approx (3\xi E_{BZ}/a_{rad} \Delta \Omega r_o^2 \delta t_{BZ} c)^{1/4}$, or 
\begin{equation}\label{eq:T_o}
k_B T'_o \sim 2.9 \ {\rm keV} \ \xi^{1/4} E_{BZ, 52}^{1/4} \delta t_{BZ, 2}^{-1/4} \Delta \Omega^{-1/4} r_{o, 12}^{-1/2}.
\end{equation}
We note that the optical depth at around this radius is large, $\tau_T \approx Y_e \kappa_T \rho r \sim 4.2 \times 10^7 \ (Y_e/0.2) M_{-2} r_{12}^{-1} \delta t_{-1}$. 
Here, $\kappa_T \sim 0.2 \ {\rm g^{-1} \ cm^2}$ is the opacity of the Thomson scattering, 
$\rho \approx M/4\pi r^2 c \delta t \sim 5.3 \times 10^{-4} \ M_{-2} r_{12}^{-2} \delta t_{-1} \ \rm g \ cm^{-3}$ is the mean density of the ejecta, 
and $M_{-2} = M/10^{-2}M_\odot$ is the isotropic mass ejection. 
Though the mass ejection, in general, is unisotropic~\citep[e.g.][]{Hotokezaka2013}, we can take into account this effect by changing $\Delta \Omega$ and $M$ appropriately. 
The fireball is accelerated due to the large internal energy, and the Lorentz factor saturates at $\Gamma \sim 4 \pi \xi E_{BZ}/ \Delta \Omega Mc^2$, or
\begin{equation}
\Gamma \sim 7.0 \ \xi E_{BZ, 52} M_{-2}^{-1} \Delta \Omega^{-1}, 
\end{equation}
which occurs at 
\begin{equation}
r_{s} \approx r_o \Gamma \sim 7.0 \times 10^{12} \ {\rm cm}  \ \xi E_{BZ, 52} M_{-2}^{-1} \Delta \Omega^{-1} r_{o, 12}.
\end{equation}
Hereafter, we simply assume $\Gamma \propto r$ in the acceleration phase.\footnote{If only a tiny fraction of Poynting energies is dissipated at $r \sim r_o$, the fireball is still magnetically dominated, and the evolution of the Lorentz factor is generally different from the above scaling; $\Gamma \propto r^{\mu}$ with $1/3 \lesssim \mu \lesssim 1$.}
For $r > r_s$, the fireball moves as a shell with a shell width $\sim r_s$ so that the temperature  decreases as $T' \propto (r/r_s)^{-2/3}$, 
while in the lateral direction $y$,  it  expands as  $y\sim r/\Gamma$ \citep[e.g.,][]{Meszaros_Rees_2000}.
Then, the fireball begins to expand almost spherically irrespective of the initial beaming angle beyond the radius given by
\begin{equation}\label{eq:r_exp}
r_{exp} \approx r_s \Gamma \sim 4.9 \times 10^{13} \ {\rm cm} \  \xi^2 E_{BZ, 52}^2 M_{-2}^{-2} \Delta \Omega^{-2} r_{o, 12}.
\end{equation}
The temperature  decreases as $T' \propto (r/r_{exp})^{-1}$  for $r  >  r_{exp}$.

The  photospheric emission from the fireball can be expected around the photospheric radius, 
\begin{equation}\label{eq:r_ph}
r_{ph}  \approx \left(\frac{Y_e \kappa_T M}{4 \pi} \right)^{1/2} \sim 2.5 \times 10^{14} \ {\rm cm} \ (Y_e/0.2)^{1/2} M_{-2}^{1/2}. 
\end{equation}
The temperature of the fireball in the comoving frame evolves as $T' \propto (r/r_o)^{-1}$ for $r_o < r < r_s$, $T' \propto (r/r_s)^{-2/3}$ for $r_s < r < r_{exp}$, and $T' \propto (r/r_{exp})^{-1}$ for $r_{exp} < r$. 
 From Eqs. (\ref{eq:r_exp}) and (\ref{eq:r_ph}),  $r_{ph} > r_{exp}$ is satisfied if $M$ is larger than the critical mass  given by 
\begin{equation}
M_{-2} > 0.76 \ (Y_e/0.2)^{-1/5} \xi^{4/5} E_{BZ, 52}^{4/5} \Delta \Omega^{-4/5} r_{o, 12}^{2/5}.
\end{equation}
In the case of {\it dirty} fireballs with $M_{-2}\sim 1$, $r_{ph} > r_{exp}$ is expected.
The temperature at the photospheric radius becomes $T'_{ph} \approx T'_o (r_s/r_o)^{-1} (r_{exp}/r_s)^{-2/3} (r_{ph}/r_{exp})^{-1}$, which is expressed as
\begin{eqnarray}
k_BT'_{ph} &\sim& 22 \ {\rm eV} \ (Y_e/0.2)^{-1/2} \notag \\
&\times& \xi^{7/12} E_{BZ, 52}^{7/12} M_{-2}^{-5/6} \delta t_{BZ, 2}^{-1/4} \Delta \Omega^{-7/12} r_{o, 12}^{1/2},
\end{eqnarray}
and the peak photon energy of the resultant photospheric emission can be estimated as $\varepsilon_{peak} \approx 2.83 k_B T'_{ph} \Gamma/(1+z)$, which is given by
\begin{eqnarray}\label{eq:ee_peak_dirty}
\varepsilon_{peak} &\sim& 0.40 \ {\rm keV} \ (1+z)^{-1} (Y_e/0.2)^{-1/2} \notag \\
&\times& \xi^{19/12} E_{BZ, 52}^{19/12} M_{-2}^{-11/6} \delta t_{BZ, 2}^{-1/4} \Delta \Omega^{-19/12} r_{o, 12}^{1/2}.
\end{eqnarray}
The peak intensity in the comoving frame can be approximated as $I'_{peak} \approx 2(\nu'_{peak}/c)^2 \times 2.83 k_B T'_{ph} \times \exp(-2.83)$, 
where $\nu'_{peak} \approx 2.83 k_B T'_{ph}/h$ with the Planck constant $h = 6.62 \times 10^{-27} \ \rm erg \ s$.\footnote{Note that here $h$ is not the half thickness of the disk but the Planck constant. Also we use $\beta$ as the spectral index in this section.}
In the observer frame, using the fact that $I_\nu/\nu^3$ is Lorentz invariant~\citep[e.g.,][]{Rybicki1979}, 
the corresponding energy flux is given by $F_{peak} \approx (\pi/\Gamma^2) \times (r_{ph}/d_L)^2 \times \Gamma^3 \times I'_{peak} \times (1+z)$ 
with $d_L$  and  $1/\Gamma^2$ being the luminosity distance and the relativistic beaming effect, respectively, as
\begin{eqnarray}\label{eq:F_peak_dirty}
F_{peak} &\approx& 0.37\times \frac{(1+z)^4}{\Gamma^2}\left( \frac{r_{ph}}{d_L}\right)^2 \frac{\varepsilon_{peak}^3}{h^2c^2} \notag \\
&\sim& 1.1 \times 10^{-23} \ {\rm erg \ cm^{-2} \ s^{-1} \ Hz^{-1}} \ (1+z) \notag \\
&\times& \ \left(\frac{d_L}{200 \ \rm Mpc}\right)^{-2} \ (Y_e/0.1)^{-1/2} \notag \\
&\times& \ \xi^{11/4} E_{BZ, 52}^{11/4} M_{-2}^{-5/2} \delta t_{BZ, 2}^{-3/4} \Delta \Omega^{-11/4} r_{o, 12}^{3/2},
\end{eqnarray}
\begin{eqnarray}\label{eq:ee_flux_dirty}
\nu_{peak}F_{peak} &\sim& 1.2 \times 10^{-6} \ {\rm erg \ cm^{-2} \ s^{-1} } \notag \\
&\times& \ \left(\frac{d_L}{200 \ \rm Mpc}\right)^{-2} \ (Y_e/0.1)^{-1} \notag \\
&\times& \ \xi^{13/3} E_{BZ, 52}^{13/3} M_{-2}^{-13/3} \delta t_{BZ, 2}^{-1} \Delta \Omega^{-13/3} r_{o, 12}^2.
\end{eqnarray}
In general, the observed duration of the photospheric emission is given by
\begin{equation}\label{eq:t_dur}
t_{dur} \approx (1+z) \times \max[r_{ph}/c\Gamma^2, \delta t_{BZ}].
\end{equation}
In the case of dirty fireballs, $r_{ph} > r_{exp} \sim c\delta t_{BZ}\Gamma^2$ (see Eqs. \ref{eq:r_o} and \ref{eq:r_exp}), and thus the first term on the right-hand side of the above equation always larger than the second one. Then,
\begin{equation}\label{eq:t_dur_dirty}
t_{dur} \sim 170 \ {\rm s} \ (1+z) (Y_e/0.2)^{1/2} \xi^{-2} E_{BZ, 52}^{-2} M_{-2}^{5/2} \Delta \Omega^2. 
\end{equation}
We remark that EE durations of $t_ {dur} > 100 \ \rm s$ do not necessarily require durations of the BZ jet of $\delta t_{BZ} > 100 \ \rm s$ 
if the fireball is dirty and the duration is determined by Eq. (\ref{eq:t_dur_dirty}). 
For a fiducial parameter set ($\xi=1, E_{BZ, 52}=1, M_{-2}=1, \delta t_{BZ, 2}=1, \Delta \Omega=1$), 
the emission is characterized by $\varepsilon_{peak} \sim 0.42 \ {\rm keV}$, $\nu_{peak}F_{peak} \sim 1.2 \times 10^{-6}\ {\rm erg \ cm^{-2} \ s^{-1} }$, and $t_{dur} \sim 180 \ \rm s$ from $d_L = 200 \ \rm Mpc$, 
and $\varepsilon_{peak} \sim 0.29 \ {\rm keV}$, $\nu_{peak}F_{peak} \sim 6.0 \times 10^{-9}\ {\rm erg \ cm^{-2} \ s^{-1} }$, and $t_{dur} \sim 260 \ \rm s$ from $z = 0.5$. 

\begin{figure}
\includegraphics[width = 150mm]{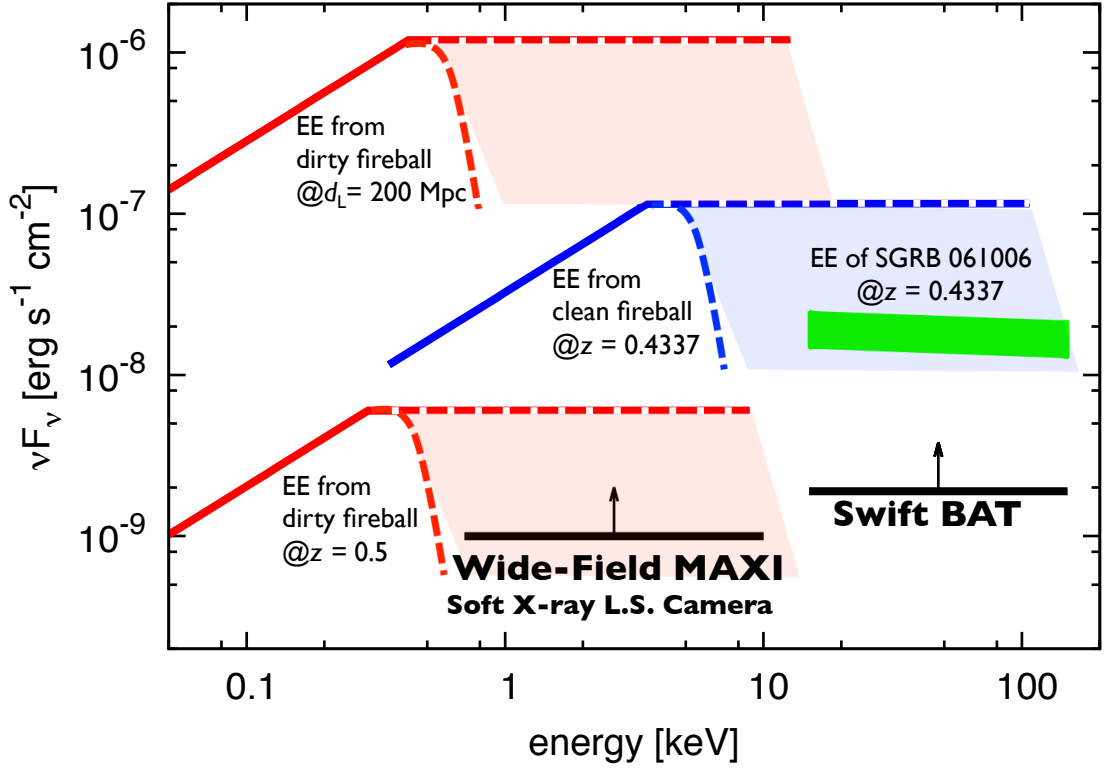}
\caption{Extended emissions (EEs) of short gamma-ray bursts (SGRBs) based on our model and their detectability. 
The red lines show the anticipated fluxes of the dissipative photospheric emissions from the dirty fireballs (Eqs. \ref{eq:ee_peak_dirty} and \ref{eq:ee_flux_dirty}) 
at $d_L = 200 \ \rm Mpc$ ($z = 0.046$) and  $d_L = 2.8 \ \rm Gpc$ (z = 0.5), 
and the blue line shows the clean fireball case (Eqs. \ref{eq:ee_peak_clean} and \ref{eq:ee_flux_clean}) at $d_L = 2.4 \ \rm Gpc$ ($z = 0.4337$). 
The shaded regions represent the uncertainties of the subphotospheric dissipations. 
If the dissipations are weak, the spectra are quasi-thermal, and if strong, the high energy tails can extend up to $\varepsilon_{cut} \lesssim 30 \times \varepsilon_{peak}$ with a photo index of $\lesssim 2$.   
The black lines with the upward arrow show the detection thresholds of Wide-Field MAXI ($0.7\mbox{-}10 \ \rm keV$) and Swift BAT ($15\mbox{-}150 \ \rm keV$).
The green thick line shows the observed flux spectrum of the EE associated with SGRB 061006.}
\label{fig:ee_obs}
\end{figure}

In general, the spectral shape is determined by additional dissipation processes, e.g., internal shocks or magnetic reconnections occurring in $r_o < r < r_{ph}$, 
which may slightly boost the peak energy and most likely produce a quasi-thermal spectrum,  
\begin{equation}\label{eq:F}
F \approx F_{peak} \times \left\{
\begin{array}{ll}
(\varepsilon/\varepsilon_{peak})^{2} & \mbox{for $\varepsilon < \varepsilon_{peak}$}, \\
(\varepsilon/\varepsilon_{peak})^{\beta}, & \mbox{for $\varepsilon_{peak} < \varepsilon < \varepsilon_{cut}$}.
\end{array}
\right. 
\end{equation}
For example, \cite{Pe'er2006} numerically calculated the photospheric emission from fireballs (in their cases, hot-plasma cocoons) including some dissipation processes at $r \lesssim r_{ph}$, 
and they showed that typically $\beta \sim -1$ and $\varepsilon_{cut} \sim 30\times \varepsilon_{peak}$.  
Then, for our fiducial parameters, $\varepsilon_{cut} \sim 13 \ {\rm keV}$  from $d_L = 200 \ \rm Mpc$ so that the spectrum may range from $\sim 0.4 \ {\rm keV}$ to $\sim 13\ {\rm keV}$. 
We plot the possible $ \nu F_\nu $ spectra in Fig. \ref{fig:ee_obs}. 
Although $\varepsilon_{peak}$ has parameter dependancies like Eq. (\ref{eq:ee_peak_dirty}), it is most likely that $\varepsilon_{cut}$ is below $15 \ \rm keV$ for $M_{-2}\sim 1$ and $\Delta \Omega \sim 1$.  
In this case, the emission energy is outside of the coverage of BATSE and {\it Swift} BAT so that such EEs have not been detected so far.
On the other hand, the soft EEs can be detected by soft X-ray survey facilities like Wide-Field MAXI~\footnote{http://spacephysics.uah.edu/gammacon/wp-content/program/program-node77.html}, 
which has a $0.1^\circ$ angular resolution.

Let us estimate the possible detection rate of the soft EEs discussed above, in particular, simultaneously with the GWs from compact binary mergers.   
\begin{eqnarray}
&& 0.2 \times 0.8 \times 0.9 \times f_{softEE} \left(\frac{\Delta \Omega_{softEE}}{4\pi}\right) {\cal R}_{GW} \notag \\
&\sim& 5.7 \ {\rm yr^{-1}} \ f_{softEE}  \left(\frac{\Delta \Omega_{softEE}}{4\pi}\right) \left(\frac{{\cal R}_{GW}}{40 \ {\rm yr^{-1}}} \right),
\end{eqnarray}
where $f_{softEE}$ is the fraction of the soft EEs in all SGRBs, 
and ${\cal R}_{GW}$ represents the NS-NS merger rate within the detection horizon of advanced LIGO, advanced VIRGO, and KAGRA, 
$d_L \sim 200 \ \rm Mpc$, which corresponds to a redshift of $z \sim 0.046$ and a comoving volume of $V_{GW} \sim 0.03 \ \rm Gpc^3$. 
We assume the standard $\Lambda$-CDM cosmology.  
Note here that $ f_{softEE} \sim 1$ can be expected given that the fraction of an EE burst is significantly larger in softer energy bands; 
$\sim 25\%$ in the {\it Swift} BAT samples ($> 15 \ \rm keV$) and $\sim 7\%$ in the BATSE samples ($> 20 \ \rm keV$). 
We take into account the sky coverage of Wide-Field MAXI $\sim 20 \ \%$ and the anticipated duty cycle, $\sim 80 \ \%$ for Wide-Field MAXI and $\sim 90 \%$ for the 2nd generation GW network. 
In the case of dirty fireballs, the beaming factor can be relatively large, e.g., $\Delta \Omega_{softEE} \sim 1$, 
and the above estimate gives $\sim 0.5 \ \rm yr^{-1}$ for ${\cal R}_{GW}\sim 40 \ {\rm yr^{-1}} $ and $ f_{softEE}\sim 1$. 
With a planned detection threshold flux of Wide-Field MAXI $\sim 1.0\times 10^{-9} \ \rm erg \ s^{-1} \ cm^{-2}$, 
such soft EEs can be detectable from $z \sim 0.5$, which corresponds to a luminosity distance of $d_L = 2.8 \ \rm Gpc$ and a comoving volume of $V_{softEE,MAXI} = 28 \ \rm Gpc^3$. 
The anticipated total detection rate can be estimated as 
\begin{equation}\label{eq:rate_soft_ee}
\sim 430 \ {\rm yr^{-1}} \ f_{softEE} \ \Delta \Omega_{softEE} \ ({\cal R}_{GW}/40 \ {\rm yr^{-1}}). 
\end{equation}

Next let us consider the observed EEs in our scenario.  
As we argued above, the dissipative photospheric emissions from dirty fireballs are too soft and too dim for {\it Swift} BAT as far as $M_{-2} \sim 1$ and $\Delta \Omega \sim 1$. 
Importantly, the typical beaming angle of the observed EEs can be estimated to be much smaller as follows.
The EEs are being detected predominantly by {\it Swift} BAT with the image trigger.
Here we set the trigger threshold as $2 \ {\rm photon \ cm^{-2}}$ in 64 s \citep{Toma_et_al_2011}.
In this case, the trigger threshold by BAT for a burst with $\beta = -1$ can be calculated as $F_{det,BAT} = 1.9 \times 10^{-9} \ {\rm erg \ s^{-1} \ cm^{-2}}$. 
For an EE with a mean luminosity of $L_{iso,15\mbox{-}150 \ \rm keV} \sim 10^{49} \ \rm erg \ s^{-1}$ (like SGRB 061006), the detection horizon also becomes $d_L = 6.6 \ \rm Gpc$, 
corresponding to $z = 1$ and $V_{EE,BAT} = 150 \ \rm Gpc^3$. 
For an effective total observation time for BAT of $T_{obs, BAT} \sim 0.8 \times 6 \ \rm yr$,  
and a sky coverage of $15 \%$, the total number of detectable EEs becomes $\sim 1.4 \times 10^5 \ ({\cal R}_{GW}/40 \ {\rm yr^{-1}})$.
Given that 14 EEs have been identified in this interval~\citep{Gompertz2013}, the fraction of EE bursts $f_{EE}$ and the beaming factor $\Delta \Omega_{EE}/4 \pi$ can be constrained as 
$f_{EE} (\Delta \Omega_{EE}/4\pi) \sim 9.7 \times 10^{-5}  \ ({\cal R}_{GW}/40 \ {\rm yr^{-1}})^{-1}$, or
\begin{equation}\label{eq:omega_ee}
\Delta \Omega_{EE} \sim 4.9 \times 10^{-3} \ \left(\frac{f_{EE}}{0.25}\right)^{-1} \left(\frac{{\cal R}_{GW}}{40 \ {\rm yr^{-1}}} \right)^{-1}, 
\end{equation}
which is small compared to the inferred beaming factor of the prompt spikes, $\Delta \Omega_{PS} /4\pi\sim 10^{-3}$ \citep{Fong2012}. 
We should mention that candidates of ``orphan" EEs without prompt spikes, i.e., long GRBs with $T_{90} \gtrsim 100$ whose redshifts and host galaxies are not identified, have been detected by {\it Swift} BAT. 
The detection rate of those candidates is roughly comparable to that of SGRBs with EEs\footnote{$\rm http://swift.gsfc.nasa.gov/archive/grb_table/$}. 
Given this fact, the constraint on the beaming factor (Eq. \ref{eq:omega_ee}) is a lower limit and can be larger by a factor.  
Nevertheless, the possible simultaneous detection rate of such EEs and GWs from NS-NS mergers would be quite small, 
$\approx 0.15 \times 0.8 \times 0.9 \times {\cal R}_{GW} f_{EE} (\Delta \Omega_{EE}/4\pi) \lesssim 10^{-3} \ \rm yr^{-1}$. 
Note that the estimated value is independent of the relatively uncertain binary NS merger rate. 

In the context of our picture, a smaller $\Delta \Omega$ corresponds to a relatively narrow fireball.  
We note that the mass ejection associated with the BH-NS merger is orders of magnitude smaller in the polar direction than the angle averaged one, 
which has been confirmed by numerical simulations \citep{Kyutoku2013} where the BH spin is set to be parallel to the orbital angular momentum. 
In general, it is expected that they are misaligned due to the kick velocity in the formation process of NSs. 
Although more numerical simulations using general initial conditions are needed to know what happens in BH-NS mergers, 
the simulations by \cite{Kyutoku2013} suggest that the matter along the BZ jet axis is small compared with an NS-NS binary.
As a result, one can expect that a smaller fraction of energy is dissipated in a smaller region of the pre-ejecta compared with the previous NS-NS cases 
where the outflows via the BZ process clash with denser pre-ejecta.  
Hereafter, we take $\xi \sim 10^{-4}$, $M \sim 10^{-4} \ M_\odot$, $\Delta \Omega \sim 10^{-2}$ as a fiducial value.

Such fireballs are {\it clean} in that $r_{ph} < r_s$, i.e., 
\begin{equation}\label{eq:m_crit_less}
M_{-4} < 2.3 \ (Y_e/0.2)^{-1/3} \xi_{-3}^{2/3} E_{BZ, 52}^{2/3} \Delta \Omega_{-2}^{-2/3}  \delta r_{o, 12}^{2/3}. 
\end{equation}
The Lorentz factor of such clean fireballs and the comoving temperature at the photospheric radius are $T'_{ph} \approx T'_o (r_{ph}/r_o)^{-1}$ and $\Gamma \approx r_{ph}/r_o$, respectively, 
and the peak energy of the photospheric emission $\varepsilon_{peak} \approx 2.83 k_B T'_{ph} \times \Gamma/(1+z)$ can be estimated as  
\begin{equation}\label{eq:ee_peak_clean}
\varepsilon_{peak} \sim 4.3 \ {\rm keV} \ (1+z)^{-1}  \xi_{-3}^{1/4} E_{BZ, 52}^{1/4} \delta t_{BZ, 2}^{-1/4} \Delta \Omega^{-1/4} r_{o, 12}^{-1/2}.
\end{equation}
From Eq. (\ref{eq:F_peak_dirty}), the corresponding peak flux is 
\begin{eqnarray}\label{eq:ee_flux_clean}
\nu_{peak}F_{peak} &\sim& 1.0 \times 10^{-5} \  {\rm erg \ cm^{-2} \ s^{-1}} \notag \\
&& \times \left(\frac{d_L}{200 \ \rm Mpc}\right)^{-2} \notag \\
&& \times  \xi_{-3} E_{BZ, 52} \delta t_{BZ, 2}^{-1} \Delta \Omega_{-2}^{-1}.
\end{eqnarray}
In the case of clean fireballs, the emission duration is $t_{dur} \approx (1+z) \times \delta t_{BZ}$.

For a fiducial parameter set ($\xi = 10^{-3}, E_{BZ, 52} = 1, M_{-4} = 1, \delta t_{BZ, 2} = 0.7, \Delta \Omega_{-2} = 1$), 
the emission is characterized by $\varepsilon_{peak} \sim 3.5 \ {\rm keV}$, $\nu_{peak}F_{peak} \sim 1.2 \times 10^{-7}\ {\rm erg \ cm^{-2} \ s^{-1} }$, 
and $t_{dur} \sim 100 \ \rm s$ from $z = 0.4337$ (see Fig. \ref{fig:ee_obs} where we also plot an observed flux spectrum of EE associated with SGRB 061006 at $z = 0.4337$). 
The observed EEs can be consistently interpreted as  the high energy tail of the photospheric emission from such a clean fireball. 
To test this scenario, simultaneous detections in the soft X-ray band ($< 10 \ \rm keV$) are crucial. 

\section{Discussions}
As for the EEs from fireballs, we need to model, e.g., the subphotospheric dissipation processes in detail to predict the spectra more precisely.  
Nevertheless, a key message here is that the typical energy of EEs is likely to be in soft X-ray bands. 
In our scenario, this is essentially due to the relatively large launching radii of the fireball, $r_o \gtrsim 10^{11} \ \rm cm$ (Eq. \ref{eq:r_o}) 
compared with that of the conventional GRB fireball, $r_o \lesssim 10^8 \ \rm cm$, which results in a lower initial temperature (Eq. \ref{eq:T_o}). 
The importance of soft X-ray bands has been implied from the observed soft photon index of EEs and the increase of the fraction of SGRB with EEs in softer bands.  
We strongly encourage soft X-ray survey facilities like Wide-Field MAXI, 
which can provide a useful electromagnetic counterpart to GWs from compact binary mergers with an angular resolution of $\sim 0.1^\circ$.
Such EE counterparts are also important in terms of time domain astronomy since they would be observed only $\sim 1 \ \rm s$ after the mergers.   
If a larger detection rate as Eq. (\ref{eq:rate_soft_ee}) is realized, 
a statistical technique using a stacking approach might also be possible for the detection of GWs, with the aid of soft EE counterparts.

 In our scenario, an EE duration of a few $10 \ \rm s$ is attributed to a relatively small disk viscosity of $\alpha \lesssim 0.01$ and the effect of the disk spreading during the accretion.
On the other hand, such a relatively long duration also may be realized if the disk accretion is suspended, but still the rotation energy of the BH is extracted 
via the interaction between the BH and the disk magnetosphere\citep{Putten2003}.

 So far, we have focused on the emission mechanism of EEs, and not discussed that of the initial spike. 
In our picture, the initial spike can be provided by the initial inhomogeneity in the ejecta. 
If there is a direction with low column density, either the  BZ jet, the neutrino-antineutrino pair annihilation jet, or the magnetic tower jet would cause the initial spike. 
Whatever origin of the initial spike is, our point here is that the major component of SGRBs can be the EEs in terms of the energetics. 
If a half opening angle of the outflow responsible for the initial spike is $\sim 0.1$, the total energy of the initial spike can be two to three orders of magnitude smaller than that of the EEs.  
Relatively soft EEs without initial spikes might already have been detected by e.g., Swift, but misidentified as other types of events. 
One might think that soft EEs should have already been detected by MAXI,~\footnote{http://maxi.riken.jp/top/}
though the rapid sky sweeping ($\sim 4\pi/90 \ \rm minutes$) makes it difficult to identify the $\sim 100 \ \rm s$ emissions.
Our scenario can be more clearly tested by future soft X-ray observations.

We note that the pressure in Eq.(29) is a factor of five larger than that estimated by the
general relativistic calculation by \cite{Chen2007}, which is due to our Newtonian treatment.
This causes a factor of five overestimate of the BZ power through Eq.(44). 
To take into account this fact as well as the difference between our Newtonian dynamics and the general relativistic one,
we add  a new phenomenological parameter $\xi_{\rm B}$ in Eq.(44) as
\begin{equation}
\frac{B^2}{8\pi} =\xi_{\rm B} ~p_{\rm ISCO},
\end{equation}
 Fig. 5 shows the total energy of  a BZ jet as a function of 
 $\alpha$ for $m_d^0 = 0.1 \ M_\odot$, $\beta_0 = 2$, $M_{BH}= 3 \ M_\odot$, and $q=0.5$ for three values of 
 $\xi_{\rm B}$ .
The total energy is obtained by integrating the BZ luminosity over time with a the mass accretion rate that is larger than the ignition rate of neutrino cooling,
which is determined by \cite{Chen2007} for $q=0.95$ (Eq. 42 of their paper). 
The dashed lines show the rotational energy of a BH with $q=0.5$ and $q=0.95$, respectively.  
For low $\alpha$, the figure shows that either the effect of back reaction is needed,  $\xi_{\rm B}$ is small or the disk accretion is suspended for a while as discussed  in previous paragraphs in this paper.  
Therefore it is urgent to undertake more precise analysis for more reliable quantitative predictions. 
Nevertheless, qualitatively our Newtonian model of EEs presented in this paper is worthwhile.

In the present paper, we treated the accretion disk in Newtonian gravity for a constant accretion rate at first, then required a finite total disk mass and the total angular momentum. 
Then, we solved the evolution of the size of the disk and the accretion rate. 
This treatment should be regarded as a crude approximation, and we should compare our results with the numerical simulations. 
Unfortunately, we could not find numerical simulations of the evolution of a finite sized accretion disk either in Kerr or Schwarzschild metric in the literature. 
One of the basic problems here is the relativistic treatment of the $\alpha$ viscosity. 
A simple generalization of the Navier Stokes equation as the one in the text book, for example, {\it Fluid mechanics} (Chapter 127) by Landau \& Lifshitz violates the causality \citep{Israel1976}, 
i.e., the information propagates with a speed faster than the light velocity. 
The reason is simple and clear. 
Let us consider the non-relativistic one-dimensional diffusion equation for some quantity $Q(t,x)$, 
\begin{equation}
\frac{\partial Q}{\partial t}=D\frac{\partial^2 Q}{\partial x^2},
\end{equation}
where $D$ is the diffusion constant.  
If we set the delta function source as
\begin{equation}
\frac{\partial^2 Q}{\partial x^2}-D^{-1}\frac{\partial Q}{\partial t}=-C\delta(x)\delta(t),
\end{equation}
 the solution for $t >0$ is expressed as
 \begin{equation}
Q=C\sqrt{\frac{1}{4\pi Dt}}\exp\left({-\frac{x^2}{4Dt}}\right).
\end{equation} 
The above solution clearly shows that the initial disturbance at $t = 0$ and $x = 0$ propagates to any $x$ even for any very small value of $t  > 0$, 
which means the causality is violated, i.e, the information propagates with infinite speed.  
The simple rule of changing the non-relativistic equation into a general relativistic one such as changing the derivative to the covariant derivative and using the projection of the tensor 
with $p^{\mu\nu}\equiv g^{\mu\nu}+u^\mu u^\nu$ does not help to guarantee the causality. 
We need to add  several new terms with undetermined parameters to the basic equations such as done by \cite{Israel1979}.
If we can start from the general relativistic Boltzmann equation, there is no problem as for the causality in principle. 
However, in practice, we should treat the distribution function that depends on three coordinates and three momenta, but it is beyond the ability of the present computer power to simulate such a problem. 
One may  think that the general relativistic resistive MHD simulations in three dimension are enough to solve this problem since the causality is not violated in such a system. 
However, the Boltzmann equation should be solved anyway since the gyration radius of the proton is typically comparable to the mean free path of p-n collision at around the ISCO.

\section{Acknowledgements}
The authors thank Kunihito Ioka, Tsvi Piran, Norita Kawanaka, Kenta Hotokezaka, Hiroki Nagakura, Koutaro Kyutoku, Peter Meszaros, Peter Veres, 
Shuichiro Inutsuka, Makoto Takamoto, and Sanemichi Takahashi for useful comments. 
This work is supported in part by the Grant-in-Aid from the Ministry of Education,
Culture, Sports, Science and Technology (MEXT) of Japan, No.23540305 (T.N.), No.24103006 (T.N.), No.23840023 (Y.S.) and JSPS fellowship (K.K.).

\begin{figure}
\includegraphics[width = 150mm]{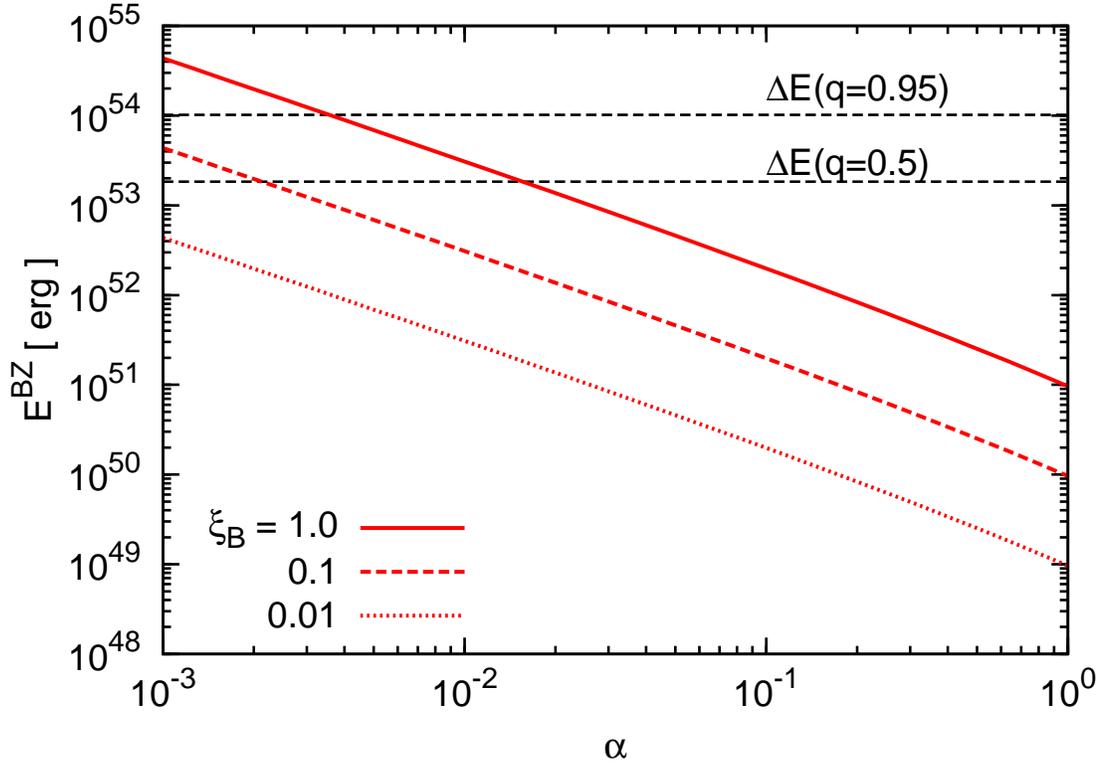}
\caption{The total BZ energy as a function of $\alpha$ for $m_d^0 = 0.1 \
M_\odot$, $\beta_0 = 2$, $M_{BH}= 3 \ M_\odot$, and $q=0.5$.
The total energy is obtained by integrating the BZ luminosity over time
during the mass accretion rate is larger than the ignition rate of
neutrino cooling, which is determined by \cite{Chen2007} for $q=0.95$ (Eq. 42 of their
paper). The red lines show the results for the different efficiency parameters
$\xi_B$ in Eq. (71): $\xi_B=1.0$ for the solid, $\xi_B=0.1$ for the
dashed, and $\xi_B=0.01$ for the dotted lines, respectively.
The dashed black lines show the rotational energy of a BH with $q=0.5$
and $q=0.95$, respectively.  For low $\alpha$, the figure shows that
either the effect of back reaction is needed, $\xi_{\rm B}$ is small or  the disk accretion is suspended for a while
as discussed  in the text.}
\label{fig:Etotal}
\end{figure}

\end{document}